\documentclass[aps,twocolumn,prl,groupedaddress,superscriptaddress,amsmath,amssymb,amsthm]{revtex4}
\usepackage{hyperref,bbm,times}
\usepackage[T1]{fontenc}
\usepackage{braket}
\usepackage{amsthm}
\usepackage{epsfig}
\usepackage{color}
\usepackage{graphicx}
\usepackage{dcolumn}
\usepackage{bm}
\usepackage[caption=false]{subfig}

\providecommand{\openone}{\leavevmode\hbox{\small1\kern-3.8pt\normalsize1}}

\usepackage{soul}

\begin{document}

\title{Quantum Information Spreading in a Disordered Quantum Walk}

\author{Farzam Nosrati}
\email{farzam.nosrati@unipa.it}
\affiliation{Dipartimento di Ingegneria, Universit\`{a} di Palermo, Viale delle Scienze, Edificio 9, 90128 Palermo, Italy}
\affiliation{INRS-EMT, 1650 Boulevard Lionel-Boulet, Varennes, Qu\'{e}bec J3X 1S2, Canada}

\author{Alessandro Laneve}
\email{alessandro.laneve@uniroma1.it}
\affiliation{Dipartimento di Fisica, Sapienza Universit`a di Roma, Piazzale Aldo Moro, 5, I-00185 Roma, Italy}

\author{Mahshid Khazaei Shadfar}
\affiliation{Dipartimento di Ingegneria, Universit\`{a} di Palermo, Viale delle Scienze, Edificio 9, 90128 Palermo, Italy}
\affiliation{INRS-EMT, 1650 Boulevard Lionel-Boulet, Varennes, Qu\'{e}bec J3X 1S2, Canada}

\author{Andrea Geraldi}
\affiliation{Dipartimento di Fisica, Sapienza Universit`a di Roma, Piazzale Aldo Moro, 5, I-00185 Roma, Italy}

\author{Kobra Mahdavipour}
\affiliation{Dipartimento di Ingegneria, Universit\`{a} di Palermo, Viale delle Scienze, Edificio 9, 90128 Palermo, Italy}
\affiliation{INRS-EMT, 1650 Boulevard Lionel-Boulet, Varennes, Qu\'{e}bec J3X 1S2, Canada}

\author{Federico Pegoraro}
\affiliation{Dipartimento di Fisica, Sapienza Universit`a di Roma, Piazzale Aldo Moro, 5, I-00185 Roma, Italy}

\author{Paolo Mataloni}
\affiliation{Dipartimento di Fisica, Sapienza Universit`a di Roma, Piazzale Aldo Moro, 5, I-00185 Roma, Italy}

\author{Rosario Lo Franco}
\affiliation{Dipartimento di Ingegneria, Universit\`{a} di Palermo, Viale delle Scienze, Edificio 6, 90128 Palermo, Italy}

\begin{abstract}

We design a quantum probing protocol using Quantum Walks to investigate the Quantum Information spreading pattern. We employ Quantum Fisher Information, as a figure of merit, to quantify extractable information about an unknown parameter encoded within the Quantum Walk evolution. Although the approach is universal, we focus on the coherent static and dynamic disorder to investigate anomalous and classical transport as well as Anderson localization. 
Our results show that a Quantum Walk can be considered as a readout device of information about defects and perturbations occurring in complex networks, both classical and quantum.

\end{abstract}

\date{\today }

\maketitle

\textit{Introduction.---} Quantum Walk (QW) is the quantum equivalent of a random walk, which benefits from quantum features such as quantum superposition, interference and entanglement \cite{Aharonov1993,venegas2012quantum,kempe2003quantum,parthasarathy1988passage}. In contrast with a classical walker, the quantum walker spreads quadratically faster in position space \cite{Aharonov1993, venegas2012quantum, kempe2003quantum, parthasarathy1988passage, hoyer2009faster}. This notion brings up a motivation to introduce algorithms for quantum computers that solve the problem exponentially faster than the best classical algorithm \cite{Childs2009Universal,douglas2008classical}. On the other hand, QW provides a powerful model to describe energy transport phenomena in heterogeneous systems either biological, in the case of photosynthesis \cite{mohseni2008environment,plenio2008dephasing}, or solid state ones, in the case of Luttinger liquids \cite{bulchandani2020superdiffusive}. Besides, nonclassical features play a substantial role in the dynamics of a quantum walker, as a sign for quantum coherence effects in biological systems \cite{engel2007evidence, lambert2013quantum}. Finally, it is worth noting that the QW model is applicable to simulate a wide range of quantum phenomena such as topological phases  \cite{obuse2011topological,kitagawa2010exploring,obuse2011topological}, neutrino oscillations \cite{mallick2017neutrino,di2016quantum}, and relativistic quantum dynamics \cite{strauch2006relativistic,chandrashekar2010relationship,Molfetta2013Quantumwalk}. 

A relevant quantity, useful to characterize the walk, is the Mean-Square Displacement (MSD) of the walker in absence of bias. The linear growth of the MSD as a function of the evolution time has become the universal identifier of what is known as normal transport. Any stochastic process that does not follow a linear growth trend with time is called anomalous. In particular, processes characterized by a superlinear growth of the MSD are usually addressed as superdiffusive \cite{metzler2000random, metzler2004restaurant, klages2008anomalous}. There are several cases in which superdiffusion settles in  transport or propagation processes, such as complex biological environments \cite{hornung2005morphogen,mohseni2008environment,plenio2008dephasing}, chaotic Hamiltonian systems \cite{balescu1995anomalous,shlesinger1993strange}, transport in disordered systems \cite{bouchaud1990anomalous,van2012exact,kim2018direct}, quantum optical systems \cite{schaufler1999keyhole}, single-molecule spectroscopy \cite{zumofen1994spectral,barkai1999distribution}. On the other hand, numerical evidence has been notified for the existence of subdiffusive transport \cite{lev2015absence, agarwal2015anomalous, vznidarivc2016diffusive, schulz2018energy, mendoza2019asymmetry}. Also, an extremely slow process of matter-wave spreading, subdiffusion transport, has been experimentally implemented in Bose-Einstein condensates \cite{lucioni2011observation}.

\begin{figure}[b!] 
\centering
\includegraphics[width=0.48\textwidth]{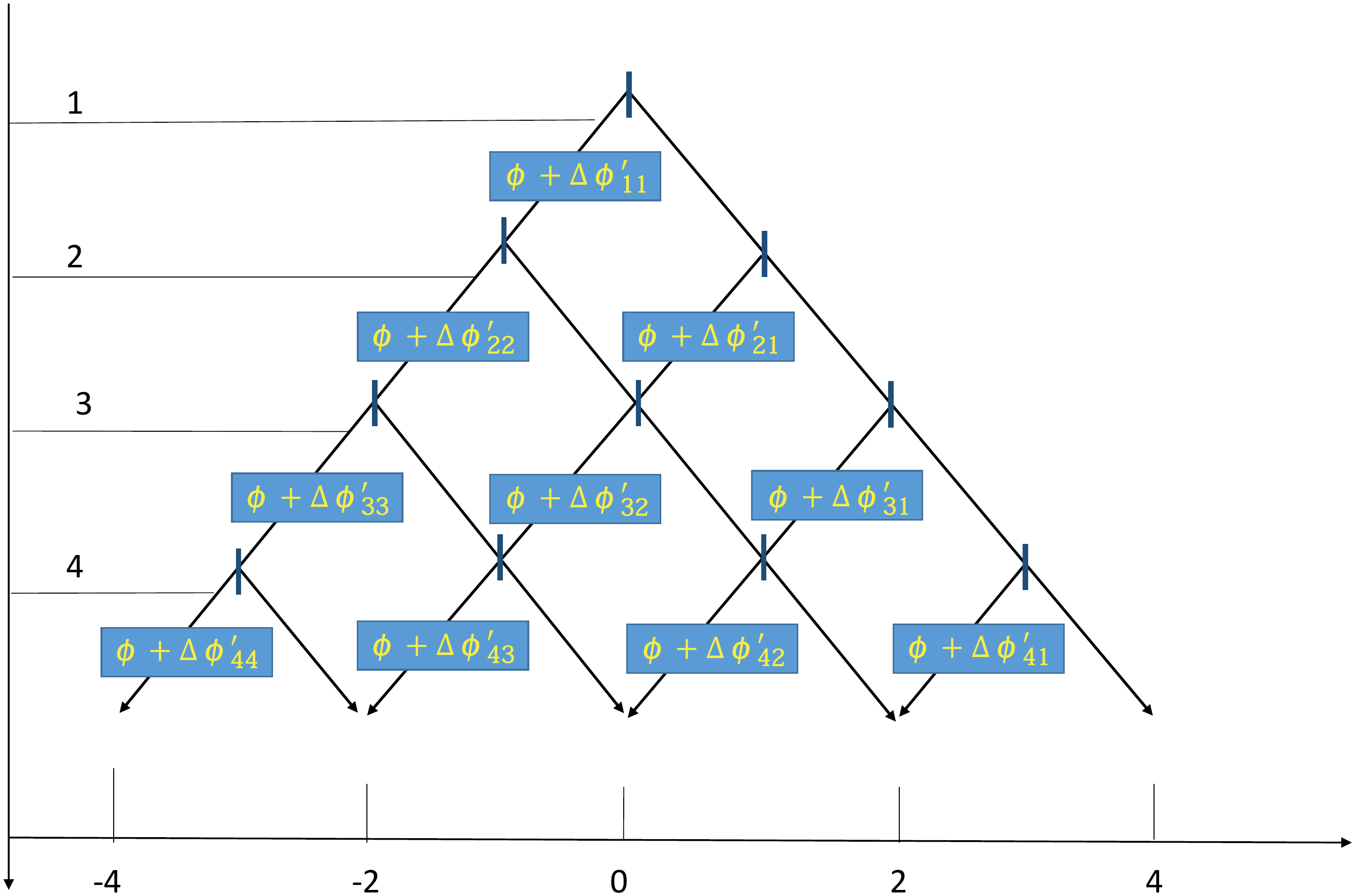}
\caption{Typical representation of a quantum probing protocol using disordered quantum walk dynamics. A phase difference $\phi$ between coin (internal) states of the walker is applied after each step, including possible fluctuations $\Delta\phi'_{i,j}$ depending on step $i$ (time) and position $j$ within the circuit. This way, static and dynamic disorder can affect the quantum walk process.}
\label{fig-repe}
\end{figure}

In recent years, there has been substantial interest in formulating QW models that can exhibit different transport behaviors with respect to the typical one, consisting in a MSD growing quadratically with time. The spreading behavior of quantum walkers can be modified by suitably tuning the evolution of a quantum walker through various types of disorder \cite{crespi2013anderson, de2014quantum, geraldi2019experimental} and decoherence effects \cite{schreiber2011decoherence}. Due to the latter ones, it has been shown that the ballistic growth of the variance changes to a superdiffusive one, reaching the diffusive spread \cite{geraldi2019experimental}. Moreover, by means of the same techniques, the subdiffusive region, between diffusive and Anderson localization regime \cite{Anderson1958localization}, can be exploited \cite{geraldi2020subdiffusion}. On the contrary, there are few reports in which a QW in presence of evolution nonlinearities avoids complete trapping in a finite region of the lattice. In these cases, the spreading has to slow down to a subdiffusive case, but it does not converge, generating a phenomenon known as delocalization of the wave packet \cite{vakulchyk2019wave}. 

\begin{figure*}[t]
\begin{center}
\subfloat[Static disorder] {\includegraphics[width=0.92\textwidth]{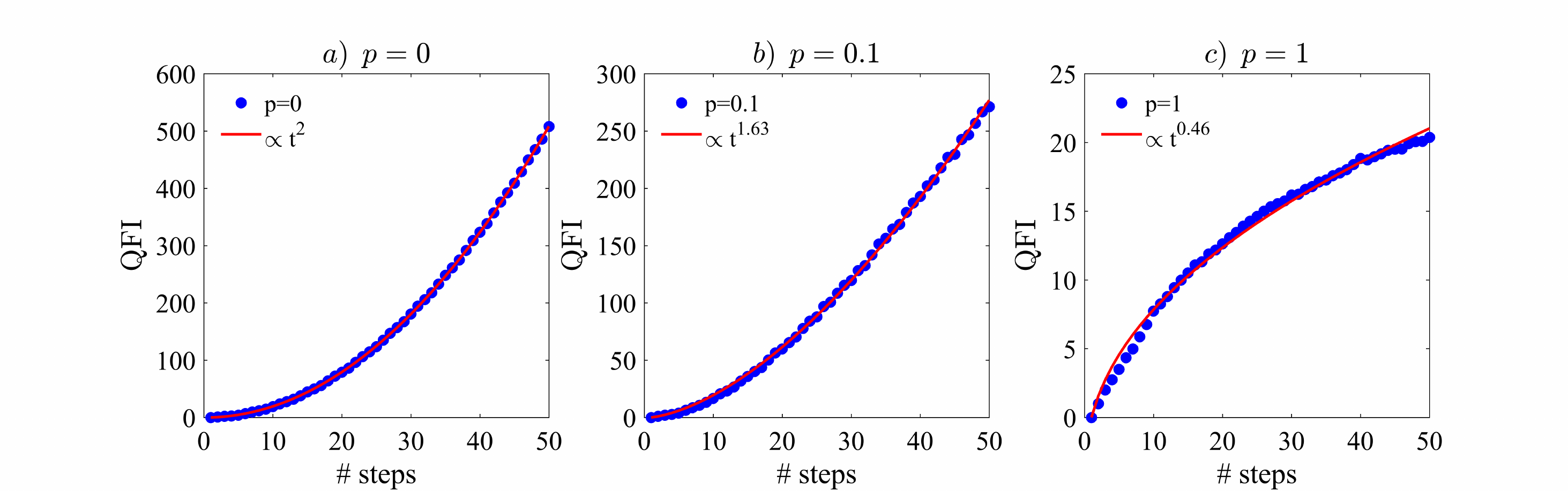}} 

\subfloat[Dynamic disorder] {\includegraphics[width=0.92\textwidth]{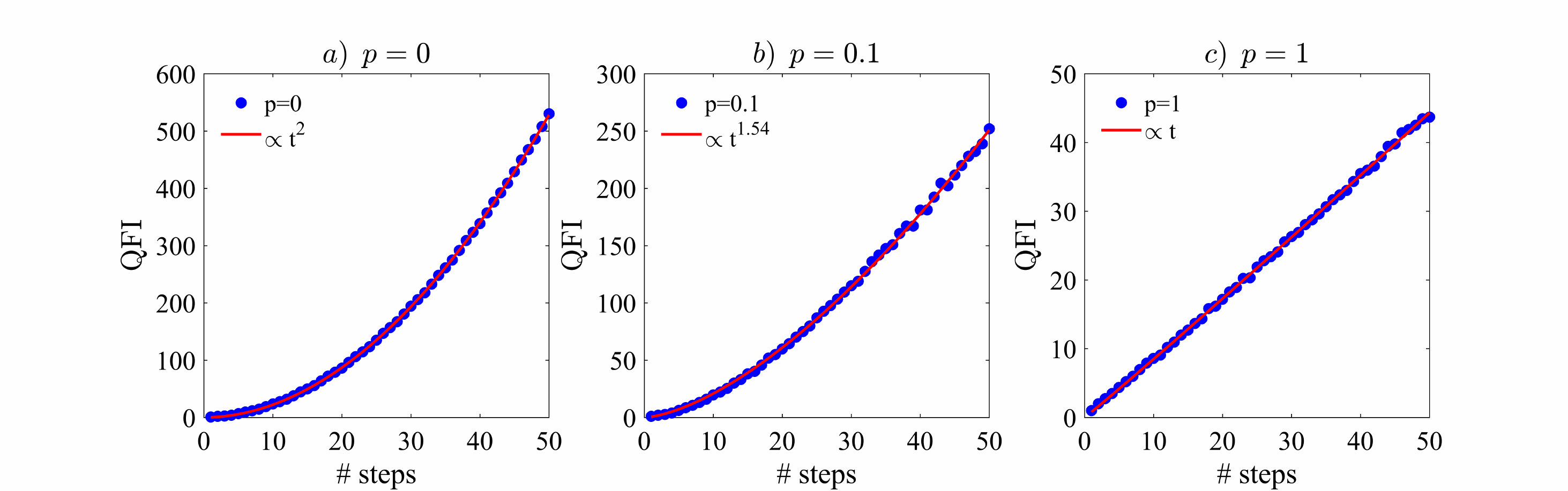}} 
\end{center}
\caption{\textbf{Quantum Fisher information for a quantum walker.} Average QFI $\mathcal{F}$ of a quantum walker versus the number of steps $t$ for static (\textbf{I}) and dynamic (\textbf{II}) disorder, at different degrees of disorder: \textbf{a)} $p=0$, \textbf{b)} $p=0.1$, \textbf{c)} $p=1$. In all plots, the initial input is $\ket{\Psi_0}=\ket{0}_{\mathbf{p}}\otimes\ket{\uparrow}_{\mathbf{c}}$.}\label{QFI-fig}
\end{figure*}

Despite the high number of studies about transport features in a QW, as far as we know, investigations in a quantum metrology fashion have remained elusive. The idea behind this work is to infer how much information about the features of the QW network can be extracted from the characteristics of the quantum walker evolution. In quantum metrology, the extractable information about an unknown parameter, such as a phase $\phi$, is usually given by the quantum Fisher information (QFI), which is also linked to the measurement accuracy of the estimation strategy \cite{giovannetti2011advances}. We take quantum Fisher information for granted to investigate the different transport regimes of information due to the properties of the quantum network. From another perspective, the growth pattern of QFI is a faithful indicator to describe defects and perturbations occurring in complex networks, both classical and quantum. We find that the disorder pattern plays a significant role in the spreading pattern of quantum information. The growth trend of QFI allows one to characterize the anomalous and normal spreading pattern, and also Anderson localization of the walker.

\textit{Theoretical model.---} The dynamics of a Quantum Walker on a one-dimensional lattice is defined on the joint Hilbert space $\mathcal{H}=\mathcal{H}_{\mathbf{p}}\otimes\mathcal{H}_{\mathbf{c}}$ of position ($\mathcal{H}_{\mathbf{p}}$) and coin ($\mathcal{H}_\mathbf{c}$) subspaces of the walker \cite{venegas2012quantum, kempe2003quantum}. The coin basis states set is $\mathcal{B}_{\mathbf{c}}=\{\uparrow,\downarrow\}$, which can be seen as an internal degree of freedom of the walker, while the position space is spanned by the discrete set $\{\ket{x}_{\mathbf{p}}\}$, which represents the sites of the lattice. The evolution of the quantum walker is determined by the coin operator $\hat{\mathcal{C}}=\frac{1}{\sqrt{2}}\left(\ket{\uparrow}_{\mathbf{c}}\bra{\uparrow}_{\mathbf{c}}+\ket{\uparrow}_{\mathbf{c}}\bra{\downarrow}_{\mathbf{c}}+\ket{\downarrow}_{\mathbf{c}}\bra{\uparrow}_{\mathbf{c}}-\ket{\downarrow}_{\mathbf{c}}\bra{\downarrow}_{\mathbf{c}}\right)$ and the shift operator $\hat{S}=\sum_x \ket{x+1}_{\mathbf{p}}\bra{x}_{\mathbf{p}}\otimes\ket{\uparrow}_{\mathbf{c}}\bra{\uparrow}_{\mathbf{c}}+\ket{x-1}_{\mathbf{p}}\bra{x}_{\mathbf{p}}\otimes\ket{\downarrow}_{\mathbf{c}}\bra{\downarrow}_{\mathbf{c}}$, which moves the walker according to the coin state. Repeated action of the unitary operator $\hat{U}=\hat{S}(\hat{\mathcal{I}}_{\mathbf{p}}\otimes\hat{\mathcal{C}})$ defines a completely ordered QW evolution, where the coin operator is uniform in both space and time. However, in order to have a general nonuniform evolution structure, we need suitable phase maps \cite{ahlbrecht2011asymptotic}. 

 \begin{figure*}[t] 
\centering
\includegraphics[width=0.85\textwidth]{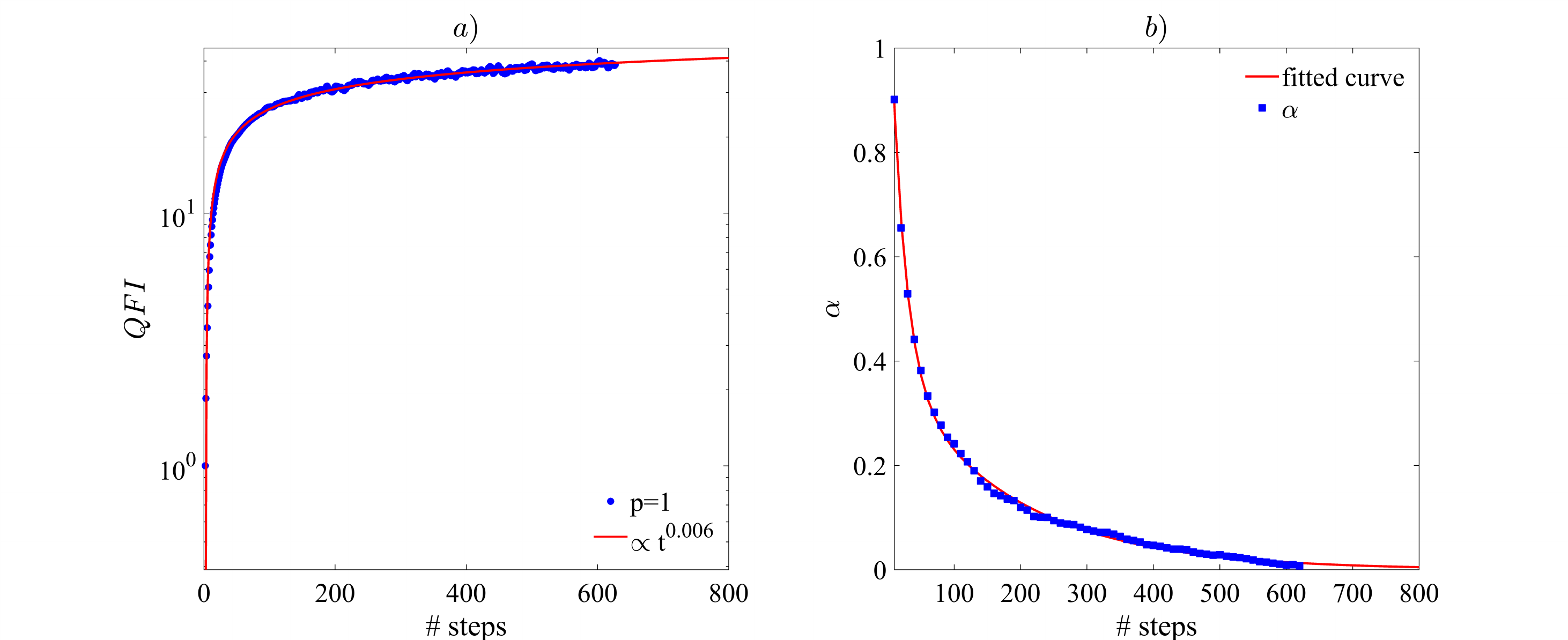}
\caption{\textbf{a)} Logarithmic scale of average QFI $\mathcal{F}$ of a quantum walker and its fitted curve $\mathcal{F}\propto t^\alpha$ as a function of the number of steps $t$ for complete static disordered $p=1$. \textbf{b)} Step-dependent coefficient $\alpha(t)$ and its fitted curve as a function of the number of steps $t$.}
\label{anderson-fig}
\end{figure*}

We design a quantum probing protocol to address the spreading behavior of a Quantum Walk and the extractable information that the walker can assess, examining the problem by quantum estimation strategy. The main aim of quantum estimation strategy is to evaluate the maximum extractable knowledge about an unknown parameter, which we call $\phi$, from repeated measurements on the probe. Typically, the parameter $\phi$ is encoded in a unitary operator \cite{giovannetti2011advances}. 
Here, we consider the case in which $\phi$ is encoded in the QW process through the unitary operator $\hat{U}(\phi)=\hat{S}(\hat{\mathcal{I}}_{\mathbf{p}}\otimes\hat{\mathcal{C}})\hat{P}$, where $\hat{P}$ is a phase-shift operator defined as
\begin{equation}
    \hat{P}=\sum_x\ket{x}_{\mathbf{p}}\bra{x}_{\mathbf{p}}\otimes\left(\ket{\downarrow}_{\mathbf{c}}\bra{\downarrow}_{\mathbf{c}}+e^{\mathbf{i}\left(\phi+\Delta\phi'(t,x)\right)}\ket{\uparrow}_{\mathbf{c}}\bra{\uparrow}_{\mathbf{c}}\right).
\end{equation}
As can be seen in Fig.~\ref{fig-repe}, the phase-shift operator $\hat{P}$ is ideally responsible for applying a phase difference $\phi$ between coin states $\ket{\downarrow}_{\mathbf{c}}$ and $\ket{\uparrow}_{\mathbf{c}}$ at each position. However, the encoding process might come with unwanted time-position-dependent fluctuations $\Delta\phi'(t,x)$ that coherently affect the evolution itself. This way, static and dynamic disorder can affect the quantum walk process (see Appendix). 

The measurement sensitivity of the phase parameter is given by the Cram\'{e}r-Rao inequality $\delta\phi\geq\delta\phi_\mathrm{min}=1/\sqrt{M\mathcal{F}_\phi}$ where $\delta\phi$ is the mean square error in the measure of parameter $\phi$ and $M$ is the number of measurements \cite{helstrom1976quantum,holevo2011probabilistic, petz1996geometries, petz2002covariance, paris2009quantum}. The QFI is $\mathcal{F}_\phi=\mathrm{Tr}[L^2\rho]$, where $\rho$ is the density operator and $L$ is the Symmetric Logarithmic Derivative (SLD) operator satisfying the equation $\mathrm{d}\rho/\mathrm{d}\phi=\{L,\rho\}/2$, with $\{\cdot,\cdot\}$ indicating the anticommutator. For a pure state $\rho^2=\rho$, the SLD operator reduces to $L=2\mathrm{d}\rho/\mathrm{d}\phi$. One can numerically obtain the QFI using the walker state at step $t$, $\ket{\Psi_t}=\hat{U}(\phi)\ket{\Psi_{t-1}}$, and its derivative with respect to parameter $\phi$, which is
\begin{equation}
    \left|\frac{\partial\Psi_{t}}{\partial\phi}\right\rangle=\frac{\partial\hat{U}(\phi)}{\partial\phi}\ket{\Psi_{t-1}}+\hat{U}(\phi)\left|\frac{\partial\Psi_{t-1}}{\partial\phi}\right\rangle.
\end{equation}

\begin{figure*}[t] 
\centering
\includegraphics[width=0.94\textwidth]{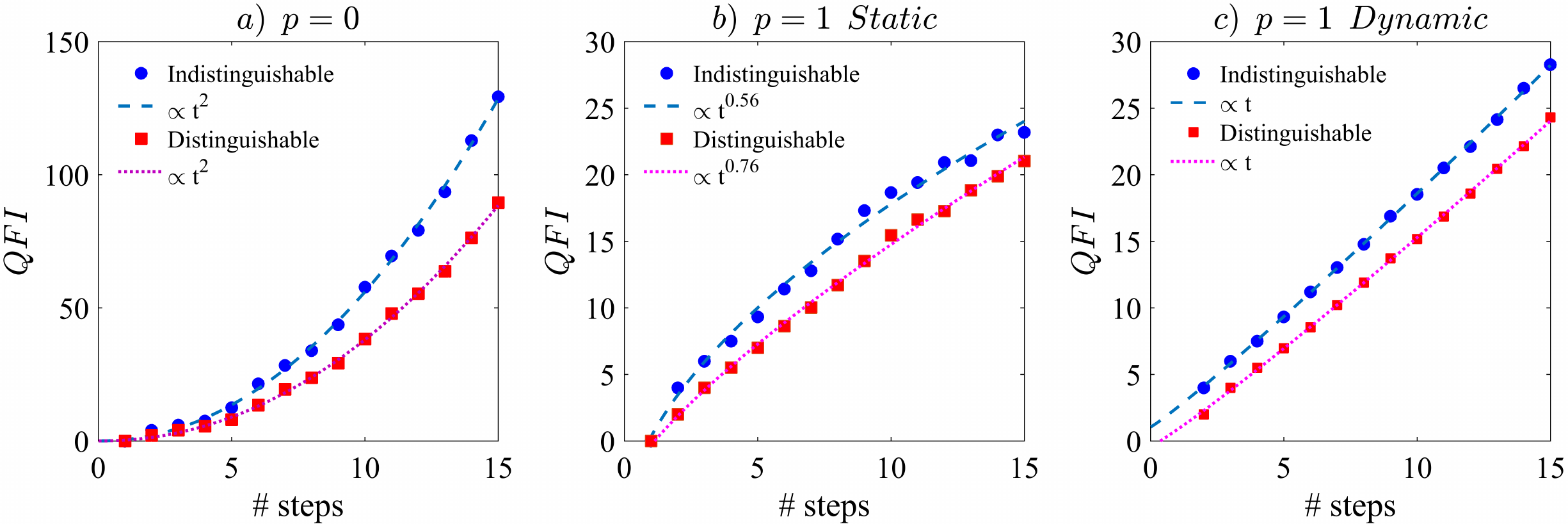}
\caption{\textbf{Quantum Fisher information for two quantum walkers.} Average QFI $\mathcal{F}$ of a quantum walker versus the number of steps $t$ for: \textbf{a)} ordered case ($p=0$), \textbf{b)} complete static disorder ($p=1$), and \textbf{c)} complete dynamic disorder ($p=1$). In all plots, the blue circles and red squares represent, respectively, indistinguishable and distinguishable two-particle inputs.}
\label{QFI2-fig}
\end{figure*}

We limit our analysis to the case in which the random fluctuations $\Delta\phi'(t,x)$ can be only $0$ or $\pi$. Afterwards, the degree of disorder $p$ is defined as the percentage of random phases that the walker experiences during the evolution. This simply means that a $p$ percentage of the random fluctuations $\Delta\phi'(t,x)$ are selected to have $\pi$ value and to be randomly distributed in time and position. By iterating over enough random phase samples, it is possible to numerically calculate the QFI in presence of a given percentage of randomness $p$. A key point in the particle evolution is the type of disorder: static disorder, where the phase fluctuations are frozen in time, or dynamic disorder, where the imposed phase can change in both time and space, as happens in the $p$-Diluted model \cite{geraldi2019experimental}. The degree of disorder $p$ is directly connected to the time evolution of the probability distribution of the quantum walker (see Appendix for the probability distribution of a quantum walker corresponding to both cases of static and dynamic disorder). 

\textit{Results.---} Let us first consider the simplest instance where a particle starts the quantum walk in the position $\ket{0}_{\mathbf{p}}$ with the $\ket{\uparrow}_{\mathbf{c}}$ coin state, that is $\ket{\Psi_0}=\ket{0}_{\mathbf{p}}\otimes\ket{\uparrow}_{\mathbf{c}}$. For both static and dynamic disorder, we simulate the behavior of the average QFI for different degrees of disorder $p$. For each value of $p$, the simulation is performed by averaging over $10^4$ different phase maps. The average QFI is plotted in Fig.~\ref{QFI-fig} as a function of the step number $t$. As a result of the power-law fitting data, we find that the average QFI is $\mathcal{F}\propto t^\alpha$ where the range of $\alpha$ is $0\leq\alpha\leq 2$ and $1\leq\alpha\leq2$ for the static disorder and the dynamic disorder, respectively. Without any disorder ($p=0$), the QFI of the QW grows quadratically in time $\mathcal{F}\propto t^2$, analogously with a ballistic transport pattern, as displayed in Fig.~\ref{QFI-fig}\textbf{a}. Also, this means that the phase variance upper-bound is proportional to the inverse of the number of steps $\delta\phi\propto t^{-1}$. For static disorder (Fig. \ref{QFI-fig} (\textbf{I})), the superdiffusive ($\alpha>1$) to subdiffusive ($\alpha<1$) transition is reachable by increasing the value of disorder $p$ out of $50$ steps. Therefore, the way information spreads in the quantum network can be determined as a result of the disorder strength. For dynamic disorder, we plot the average QFI in Fig.~\ref{QFI-fig} (\textbf{II}). Here, we use the QFI to probe the transition from the ballistic regime with $p=0$ (Fig.~\ref{QFI-fig}\textbf{a}), superdiffusive with $p=0.1$ (Fig.~\ref{QFI-fig}\textbf{b}) and diffusive one with maximum degree of disorder $p=1$ (Fig.~\ref{QFI-fig}\textbf{c}), analogous to the case of classical probe. This indicates that the output information, which is inferred trough measurements performed exclusively on the probe, shows a superdiffusive to classical transition in transport pattern. The observed fluctuation in the QFI value is due to the limited number of iterations which can be realized for each step number. Similarly to the variance of the position operator of the quantum walker \cite{schreiber2011decoherence,geraldi2019experimental,geraldi2020subdiffusion} (see also Appendix), QFI provides a simple measure to quantify the transport pattern of the walker. 

It is worth mentioning that the probing pattern depends on the number of steps that a walker takes. In the static disorder case, we find out that the QFI eventually tends to a given value. This is a clear signature of Anderson localization given by average upper-bound of QFI, as clearly seen in Fig.~\ref{anderson-fig}\textbf{a} where the growth trend declines until average QFI reaches a maximum value. As an additional clarification, the QFI is fitted with the power function $\mathcal{F}\propto t^{\alpha(t)}$, where $\alpha(t)$ is the step-dependent coefficient in a nonlinear fitting process. In Fig.~\ref{anderson-fig}\textbf{b}, $\alpha(t)$ is plotted by increasing the step number of the QW process. We observe how the growth trend depends on the step number $t$, and how it significantly declines for higher $t$. This property shows that information stops spreading within the quantum network as an indicator of particle localization.

To enrich the physics of the phenomenon, we now consider two input quantum walkers, either distinguishable or indistinguishable, in position $\ket{0}_\mathbf{p}$ with initial opposite coin states. For distinguishable particles (walkers) named 1 and 2, the input state is separable: $\ket{\Psi_0^\mathrm{s}}=\ket{0}_{\mathbf{1p}}\ket{\uparrow}_{\mathbf{1c}}\otimes\ket{0}_{\mathbf{2p}}\ket{\downarrow}_{\mathbf{2c}}$. For indistinguishable particles, the input state is physically entangled in the coin states by virtue of complete spatial mode overlap \cite{lofranco2016,plenio2014PRL}: in the no-label approach \cite{lofranco2016,compagno2018} this state is simply written as $\ket{\Psi_0}=\ket{0_\mathbf{p}\uparrow_\mathbf{c},0_\mathbf{p}\downarrow_\mathbf{c}}$, which in the first-quantization formalism with labels becomes the symmetrized state $\ket{\Psi_0^{\pm}}=\ket{0}_{\mathbf{1p}}\ket{0}_{\mathbf{2p}}\otimes\left(\frac{1}{\sqrt{2}}(\ket{\uparrow}_{\mathbf{1c}}\ket{\downarrow}_{\mathbf{2c}}\pm\ket{\downarrow}_{\mathbf{1c}}\ket{\uparrow}_{\mathbf{2c}})\right)$ ($\pm$ being for bosons and fermions, respectively). The evolution can be studied by repeatedly applying the two-particle unitary operator $\hat{U}(\phi)\otimes \hat{U}(\phi)$ to the states above.
We plot the average QFI versus the step number $t$ in Fig.~\ref{QFI2-fig} for the ordered case $p=0$ (\textbf{a}) and for the completely disordered one $p=1$, with static (\textbf{b}) and dynamic (\textbf{c}) disorder. In general, the state of two indistinguishable particles exhibits a higher value of QFI compared to the distinguishable one. This property is explained by the fact that particle indistinguishability is an enriching resource for quantum information distribution within a composite system of identical particles \cite{lofranco2018PRL,nosrati2019npj,sunetalexp}. 
Interestingly, we also notice that both input states follow the same spreading pattern. In the case of ordered case ($p=0$), the growth pattern is ballistic, while for the completely disordered one ($p=1$), the QFI follows a subdiffusive pattern (Fig.~\ref{QFI2-fig}\textbf{b}) and a classical one (Fig.~\ref{QFI2-fig}\textbf{c}) due to static and dynamic disorder, respectively.

\textit{Conclusion.---} In this Letter, we have proposed a quantum probing protocol using the quantum walk process to infer information about defects and perturbations occurring in both quantum and classical networks. This goal has been achieved by applying quantum metrology techniques to the QW process. We have exploited QFI to describe extractable information concerning an unknown phase $\phi$ that the quantum walker acquires at each step, plus random fluctuations, through the QW. Even though the framework is general, we have studied coherent static and dynamic disorder in the QW to describe the transport pattern of information about the unknown parameter $\phi$. We have found that different disorder regimes, corresponding to a disorder percentage $p$ in the QW process, lead to different spreading patterns, including ballistic, superdiffusive, classical, subdiffusive regimes, and Anderson localization. Quantum Fisher information, as a spreading pattern indicator, is independent of the number of quantum walkers, in contrast with the position variance dimension which grows accordingly to the particle number \cite{geraldi2019experimental}. 
Ultimately, our results show that QW can play the role of a readout device of information about internal characteristics of complex networks.

\begin{acknowledgments}
F.N. would like to thank Ilenia Tinnirello for supporting this research and Daniele Croce for useful discussions.
\end{acknowledgments}


\begin{thebibliography}{57}
\expandafter\ifx\csname natexlab\endcsname\relax\def\natexlab#1{#1}\fi
\expandafter\ifx\csname bibnamefont\endcsname\relax
  \def\bibnamefont#1{#1}\fi
\expandafter\ifx\csname bibfnamefont\endcsname\relax
  \def\bibfnamefont#1{#1}\fi
\expandafter\ifx\csname citenamefont\endcsname\relax
  \def\citenamefont#1{#1}\fi
\expandafter\ifx\csname url\endcsname\relax
  \def\url#1{\texttt{#1}}\fi
\expandafter\ifx\csname urlprefix\endcsname\relax\def\urlprefix{URL }\fi
\providecommand{\bibinfo}[2]{#2}
\providecommand{\eprint}[2][]{\url{#2}}

\bibitem[{\citenamefont{Aharonov et~al.}(1993)\citenamefont{Aharonov,
  Davidovich, and Zagury}}]{Aharonov1993}
\bibinfo{author}{\bibfnamefont{Y.}~\bibnamefont{Aharonov}},
  \bibinfo{author}{\bibfnamefont{L.}~\bibnamefont{Davidovich}},
  \bibnamefont{and} \bibinfo{author}{\bibfnamefont{N.}~\bibnamefont{Zagury}},
  \bibinfo{journal}{Phys. Rev. A} \textbf{\bibinfo{volume}{48}},
  \bibinfo{pages}{1687} (\bibinfo{year}{1993}).

\bibitem[{\citenamefont{Venegas-Andraca}(2012)}]{venegas2012quantum}
\bibinfo{author}{\bibfnamefont{S.~E.} \bibnamefont{Venegas-Andraca}},
  \bibinfo{journal}{Quant. Inf. Process.} \textbf{\bibinfo{volume}{11}},
  \bibinfo{pages}{1015} (\bibinfo{year}{2012}).

\bibitem[{\citenamefont{Kempe}(2003)}]{kempe2003quantum}
\bibinfo{author}{\bibfnamefont{J.}~\bibnamefont{Kempe}},
  \bibinfo{journal}{Contemp. Phys.} \textbf{\bibinfo{volume}{44}},
  \bibinfo{pages}{307} (\bibinfo{year}{2003}).

\bibitem[{\citenamefont{Parthasarathy}(1988)}]{parthasarathy1988passage}
\bibinfo{author}{\bibfnamefont{K.}~\bibnamefont{Parthasarathy}},
  \bibinfo{journal}{J. App. Prob.} pp. \bibinfo{pages}{151--166}
  (\bibinfo{year}{1988}).

\bibitem[{\citenamefont{Hoyer and Meyer}(2009)}]{hoyer2009faster}
\bibinfo{author}{\bibfnamefont{S.}~\bibnamefont{Hoyer}} \bibnamefont{and}
  \bibinfo{author}{\bibfnamefont{D.~A.} \bibnamefont{Meyer}},
  \bibinfo{journal}{Phys. Rev. A} \textbf{\bibinfo{volume}{79}},
  \bibinfo{pages}{024307} (\bibinfo{year}{2009}).

\bibitem[{\citenamefont{Childs}(2009)}]{Childs2009Universal}
\bibinfo{author}{\bibfnamefont{A.~M.} \bibnamefont{Childs}},
  \bibinfo{journal}{Phys. Rev. Lett.} \textbf{\bibinfo{volume}{102}},
  \bibinfo{pages}{180501} (\bibinfo{year}{2009}).

\bibitem[{\citenamefont{Douglas and Wang}(2008)}]{douglas2008classical}
\bibinfo{author}{\bibfnamefont{B.~L.} \bibnamefont{Douglas}} \bibnamefont{and}
  \bibinfo{author}{\bibfnamefont{J.~B.} \bibnamefont{Wang}},
  \bibinfo{journal}{J. Phys. A: Math. Theor.} \textbf{\bibinfo{volume}{41}},
  \bibinfo{pages}{075303} (\bibinfo{year}{2008}).

\bibitem[{\citenamefont{Mohseni et~al.}(2008)\citenamefont{Mohseni, Rebentrost,
  Lloyd, and Aspuru-Guzik}}]{mohseni2008environment}
\bibinfo{author}{\bibfnamefont{M.}~\bibnamefont{Mohseni}},
  \bibinfo{author}{\bibfnamefont{P.}~\bibnamefont{Rebentrost}},
  \bibinfo{author}{\bibfnamefont{S.}~\bibnamefont{Lloyd}}, \bibnamefont{and}
  \bibinfo{author}{\bibfnamefont{A.}~\bibnamefont{Aspuru-Guzik}},
  \bibinfo{journal}{J. Chem. Phys.} \textbf{\bibinfo{volume}{129}},
  \bibinfo{pages}{11B603} (\bibinfo{year}{2008}).

\bibitem[{\citenamefont{Plenio and Huelga}(2008)}]{plenio2008dephasing}
\bibinfo{author}{\bibfnamefont{M.~B.} \bibnamefont{Plenio}} \bibnamefont{and}
  \bibinfo{author}{\bibfnamefont{S.~F.} \bibnamefont{Huelga}},
  \bibinfo{journal}{New J. Phys.} \textbf{\bibinfo{volume}{10}},
  \bibinfo{pages}{113019} (\bibinfo{year}{2008}).

\bibitem[{\citenamefont{Bulchandani et~al.}(2020)\citenamefont{Bulchandani,
  Karrasch, and Moore}}]{bulchandani2020superdiffusive}
\bibinfo{author}{\bibfnamefont{V.~B.} \bibnamefont{Bulchandani}},
  \bibinfo{author}{\bibfnamefont{C.}~\bibnamefont{Karrasch}}, \bibnamefont{and}
  \bibinfo{author}{\bibfnamefont{J.~E.} \bibnamefont{Moore}},
  \bibinfo{journal}{PNAS} \textbf{\bibinfo{volume}{117}},
  \bibinfo{pages}{12713} (\bibinfo{year}{2020}).

\bibitem[{\citenamefont{Engel et~al.}(2007)\citenamefont{Engel, Calhoun, Read,
  Ahn, Man{\v{c}}al, Cheng, Blankenship, and Fleming}}]{engel2007evidence}
\bibinfo{author}{\bibfnamefont{G.~S.} \bibnamefont{Engel}},
  \bibinfo{author}{\bibfnamefont{T.~R.} \bibnamefont{Calhoun}},
  \bibinfo{author}{\bibfnamefont{E.~L.} \bibnamefont{Read}},
  \bibinfo{author}{\bibfnamefont{T.-K.} \bibnamefont{Ahn}},
  \bibinfo{author}{\bibfnamefont{T.}~\bibnamefont{Man{\v{c}}al}},
  \bibinfo{author}{\bibfnamefont{Y.-C.} \bibnamefont{Cheng}},
  \bibinfo{author}{\bibfnamefont{R.~E.} \bibnamefont{Blankenship}},
  \bibnamefont{and} \bibinfo{author}{\bibfnamefont{G.~R.}
  \bibnamefont{Fleming}}, \bibinfo{journal}{Nature}
  \textbf{\bibinfo{volume}{446}}, \bibinfo{pages}{782} (\bibinfo{year}{2007}).

\bibitem[{\citenamefont{Lambert et~al.}(2013)\citenamefont{Lambert, Chen,
  Cheng, Li, Chen, and Nori}}]{lambert2013quantum}
\bibinfo{author}{\bibfnamefont{N.}~\bibnamefont{Lambert}},
  \bibinfo{author}{\bibfnamefont{Y.-N.} \bibnamefont{Chen}},
  \bibinfo{author}{\bibfnamefont{Y.-C.} \bibnamefont{Cheng}},
  \bibinfo{author}{\bibfnamefont{C.-M.} \bibnamefont{Li}},
  \bibinfo{author}{\bibfnamefont{G.-Y.} \bibnamefont{Chen}}, \bibnamefont{and}
  \bibinfo{author}{\bibfnamefont{F.}~\bibnamefont{Nori}},
  \bibinfo{journal}{Nat. Phys.} \textbf{\bibinfo{volume}{9}},
  \bibinfo{pages}{10} (\bibinfo{year}{2013}).

\bibitem[{\citenamefont{Obuse and Kawakami}(2011)}]{obuse2011topological}
\bibinfo{author}{\bibfnamefont{H.}~\bibnamefont{Obuse}} \bibnamefont{and}
  \bibinfo{author}{\bibfnamefont{N.}~\bibnamefont{Kawakami}},
  \bibinfo{journal}{Phys. Rev. B} \textbf{\bibinfo{volume}{84}},
  \bibinfo{pages}{195139} (\bibinfo{year}{2011}).

\bibitem[{\citenamefont{Kitagawa et~al.}(2010)\citenamefont{Kitagawa, Rudner,
  Berg, and Demler}}]{kitagawa2010exploring}
\bibinfo{author}{\bibfnamefont{T.}~\bibnamefont{Kitagawa}},
  \bibinfo{author}{\bibfnamefont{M.~S.} \bibnamefont{Rudner}},
  \bibinfo{author}{\bibfnamefont{E.}~\bibnamefont{Berg}}, \bibnamefont{and}
  \bibinfo{author}{\bibfnamefont{E.}~\bibnamefont{Demler}},
  \bibinfo{journal}{Phys. Rev. A} \textbf{\bibinfo{volume}{82}},
  \bibinfo{pages}{033429} (\bibinfo{year}{2010}).

\bibitem[{\citenamefont{Mallick et~al.}(2017)\citenamefont{Mallick, Mandal, and
  Chandrashekar}}]{mallick2017neutrino}
\bibinfo{author}{\bibfnamefont{A.}~\bibnamefont{Mallick}},
  \bibinfo{author}{\bibfnamefont{S.}~\bibnamefont{Mandal}}, \bibnamefont{and}
  \bibinfo{author}{\bibfnamefont{C.}~\bibnamefont{Chandrashekar}},
  \bibinfo{journal}{Eur. Phys. J. C} \textbf{\bibinfo{volume}{77}},
  \bibinfo{pages}{1} (\bibinfo{year}{2017}).

\bibitem[{\citenamefont{Di~Molfetta and P{\'e}rez}(2016)}]{di2016quantum}
\bibinfo{author}{\bibfnamefont{G.}~\bibnamefont{Di~Molfetta}} \bibnamefont{and}
  \bibinfo{author}{\bibfnamefont{A.}~\bibnamefont{P{\'e}rez}},
  \bibinfo{journal}{New J. Phys.} \textbf{\bibinfo{volume}{18}},
  \bibinfo{pages}{103038} (\bibinfo{year}{2016}).

\bibitem[{\citenamefont{Strauch}(2006)}]{strauch2006relativistic}
\bibinfo{author}{\bibfnamefont{F.~W.} \bibnamefont{Strauch}},
  \bibinfo{journal}{Phys. Rev. A} \textbf{\bibinfo{volume}{73}},
  \bibinfo{pages}{054302} (\bibinfo{year}{2006}).

\bibitem[{\citenamefont{Chandrashekar et~al.}(2010)\citenamefont{Chandrashekar,
  Banerjee, and Srikanth}}]{chandrashekar2010relationship}
\bibinfo{author}{\bibfnamefont{C.}~\bibnamefont{Chandrashekar}},
  \bibinfo{author}{\bibfnamefont{S.}~\bibnamefont{Banerjee}}, \bibnamefont{and}
  \bibinfo{author}{\bibfnamefont{R.}~\bibnamefont{Srikanth}},
  \bibinfo{journal}{Phys. Rev. A} \textbf{\bibinfo{volume}{81}},
  \bibinfo{pages}{062340} (\bibinfo{year}{2010}).

\bibitem[{\citenamefont{Di~Molfetta et~al.}(2013)\citenamefont{Di~Molfetta,
  Brachet, and Debbasch}}]{Molfetta2013Quantumwalk}
\bibinfo{author}{\bibfnamefont{G.}~\bibnamefont{Di~Molfetta}},
  \bibinfo{author}{\bibfnamefont{M.}~\bibnamefont{Brachet}}, \bibnamefont{and}
  \bibinfo{author}{\bibfnamefont{F.}~\bibnamefont{Debbasch}},
  \bibinfo{journal}{Phys. Rev. A} \textbf{\bibinfo{volume}{88}},
  \bibinfo{pages}{042301} (\bibinfo{year}{2013}).

\bibitem[{\citenamefont{Metzler and Klafter}(2000)}]{metzler2000random}
\bibinfo{author}{\bibfnamefont{R.}~\bibnamefont{Metzler}} \bibnamefont{and}
  \bibinfo{author}{\bibfnamefont{J.}~\bibnamefont{Klafter}},
  \bibinfo{journal}{Phys. Rep.} \textbf{\bibinfo{volume}{339}},
  \bibinfo{pages}{1} (\bibinfo{year}{2000}).

\bibitem[{\citenamefont{Metzler and Klafter}(2004)}]{metzler2004restaurant}
\bibinfo{author}{\bibfnamefont{R.}~\bibnamefont{Metzler}} \bibnamefont{and}
  \bibinfo{author}{\bibfnamefont{J.}~\bibnamefont{Klafter}},
  \bibinfo{journal}{J. Phys. A: Math. Theor.} \textbf{\bibinfo{volume}{37}},
  \bibinfo{pages}{R161} (\bibinfo{year}{2004}).

\bibitem[{\citenamefont{Klages et~al.}(2008)\citenamefont{Klages, Radons, and
  Sokolov}}]{klages2008anomalous}
\bibinfo{author}{\bibfnamefont{R.}~\bibnamefont{Klages}},
  \bibinfo{author}{\bibfnamefont{G.}~\bibnamefont{Radons}}, \bibnamefont{and}
  \bibinfo{author}{\bibfnamefont{I.~M.} \bibnamefont{Sokolov}},
  \emph{\bibinfo{title}{Anomalous transport: foundations and applications}}
  (\bibinfo{publisher}{John Wiley \& Sons}, \bibinfo{year}{2008}).

\bibitem[{\citenamefont{Hornung et~al.}(2005)\citenamefont{Hornung, Berkowitz,
  and Barkai}}]{hornung2005morphogen}
\bibinfo{author}{\bibfnamefont{G.}~\bibnamefont{Hornung}},
  \bibinfo{author}{\bibfnamefont{B.}~\bibnamefont{Berkowitz}},
  \bibnamefont{and} \bibinfo{author}{\bibfnamefont{N.}~\bibnamefont{Barkai}},
  \bibinfo{journal}{Phys. Rev. E} \textbf{\bibinfo{volume}{72}},
  \bibinfo{pages}{041916} (\bibinfo{year}{2005}).

\bibitem[{\citenamefont{Balescu}(1995)}]{balescu1995anomalous}
\bibinfo{author}{\bibfnamefont{R.}~\bibnamefont{Balescu}},
  \bibinfo{journal}{Phys. Rev. E} \textbf{\bibinfo{volume}{51}},
  \bibinfo{pages}{4807} (\bibinfo{year}{1995}).

\bibitem[{\citenamefont{Shlesinger et~al.}(1993)\citenamefont{Shlesinger,
  Zaslavsky, and Klafter}}]{shlesinger1993strange}
\bibinfo{author}{\bibfnamefont{M.~F.} \bibnamefont{Shlesinger}},
  \bibinfo{author}{\bibfnamefont{G.~M.} \bibnamefont{Zaslavsky}},
  \bibnamefont{and} \bibinfo{author}{\bibfnamefont{J.}~\bibnamefont{Klafter}},
  \bibinfo{journal}{Nature} \textbf{\bibinfo{volume}{363}}, \bibinfo{pages}{31}
  (\bibinfo{year}{1993}).

\bibitem[{\citenamefont{Bouchaud and Georges}(1990)}]{bouchaud1990anomalous}
\bibinfo{author}{\bibfnamefont{J.-P.} \bibnamefont{Bouchaud}} \bibnamefont{and}
  \bibinfo{author}{\bibfnamefont{A.}~\bibnamefont{Georges}},
  \bibinfo{journal}{Phys. Rep.} \textbf{\bibinfo{volume}{195}},
  \bibinfo{pages}{127} (\bibinfo{year}{1990}).

\bibitem[{\citenamefont{van Beijeren}(2012)}]{van2012exact}
\bibinfo{author}{\bibfnamefont{H.}~\bibnamefont{van Beijeren}},
  \bibinfo{journal}{Phys. Rev. Lett.} \textbf{\bibinfo{volume}{108}},
  \bibinfo{pages}{180601} (\bibinfo{year}{2012}).

\bibitem[{\citenamefont{Kim et~al.}(2018)\citenamefont{Kim, Mart{\'\i}nez,
  Phenisee, Kevrekidis, Porter, and Yang}}]{kim2018direct}
\bibinfo{author}{\bibfnamefont{E.}~\bibnamefont{Kim}},
  \bibinfo{author}{\bibfnamefont{A.~J.} \bibnamefont{Mart{\'\i}nez}},
  \bibinfo{author}{\bibfnamefont{S.~E.} \bibnamefont{Phenisee}},
  \bibinfo{author}{\bibfnamefont{P.}~\bibnamefont{Kevrekidis}},
  \bibinfo{author}{\bibfnamefont{M.~A.} \bibnamefont{Porter}},
  \bibnamefont{and} \bibinfo{author}{\bibfnamefont{J.}~\bibnamefont{Yang}},
  \bibinfo{journal}{Nat. Comm.} \textbf{\bibinfo{volume}{9}},
  \bibinfo{pages}{1} (\bibinfo{year}{2018}).

\bibitem[{\citenamefont{Schaufler et~al.}(1999)\citenamefont{Schaufler,
  Schleich, and Yakovlev}}]{schaufler1999keyhole}
\bibinfo{author}{\bibfnamefont{S.}~\bibnamefont{Schaufler}},
  \bibinfo{author}{\bibfnamefont{W.~P.} \bibnamefont{Schleich}},
  \bibnamefont{and} \bibinfo{author}{\bibfnamefont{V.~P.}
  \bibnamefont{Yakovlev}}, \bibinfo{journal}{Phys. Rev. Lett.}
  \textbf{\bibinfo{volume}{83}}, \bibinfo{pages}{3162} (\bibinfo{year}{1999}).

\bibitem[{\citenamefont{Zumofen and Klafter}(1994)}]{zumofen1994spectral}
\bibinfo{author}{\bibfnamefont{G.}~\bibnamefont{Zumofen}} \bibnamefont{and}
  \bibinfo{author}{\bibfnamefont{J.}~\bibnamefont{Klafter}},
  \bibinfo{journal}{Chem. Phys. Lett.} \textbf{\bibinfo{volume}{219}},
  \bibinfo{pages}{303} (\bibinfo{year}{1994}).

\bibitem[{\citenamefont{Barkai and Silbey}(1999)}]{barkai1999distribution}
\bibinfo{author}{\bibfnamefont{E.}~\bibnamefont{Barkai}} \bibnamefont{and}
  \bibinfo{author}{\bibfnamefont{R.}~\bibnamefont{Silbey}},
  \bibinfo{journal}{Chem. Phys. Lett.} \textbf{\bibinfo{volume}{310}},
  \bibinfo{pages}{287} (\bibinfo{year}{1999}).

\bibitem[{\citenamefont{Lev et~al.}(2015)\citenamefont{Lev, Cohen, and
  Reichman}}]{lev2015absence}
\bibinfo{author}{\bibfnamefont{Y.~B.} \bibnamefont{Lev}},
  \bibinfo{author}{\bibfnamefont{G.}~\bibnamefont{Cohen}}, \bibnamefont{and}
  \bibinfo{author}{\bibfnamefont{D.~R.} \bibnamefont{Reichman}},
  \bibinfo{journal}{Phys. Rev. Lett.} \textbf{\bibinfo{volume}{114}},
  \bibinfo{pages}{100601} (\bibinfo{year}{2015}).

\bibitem[{\citenamefont{Agarwal et~al.}(2015)\citenamefont{Agarwal,
  Gopalakrishnan, Knap, M{\"u}ller, and Demler}}]{agarwal2015anomalous}
\bibinfo{author}{\bibfnamefont{K.}~\bibnamefont{Agarwal}},
  \bibinfo{author}{\bibfnamefont{S.}~\bibnamefont{Gopalakrishnan}},
  \bibinfo{author}{\bibfnamefont{M.}~\bibnamefont{Knap}},
  \bibinfo{author}{\bibfnamefont{M.}~\bibnamefont{M{\"u}ller}},
  \bibnamefont{and} \bibinfo{author}{\bibfnamefont{E.}~\bibnamefont{Demler}},
  \bibinfo{journal}{Phys. Rev. Lett.} \textbf{\bibinfo{volume}{114}},
  \bibinfo{pages}{160401} (\bibinfo{year}{2015}).

\bibitem[{\citenamefont{{\v{Z}}nidari{\v{c}}
  et~al.}(2016)\citenamefont{{\v{Z}}nidari{\v{c}}, Scardicchio, and
  Varma}}]{vznidarivc2016diffusive}
\bibinfo{author}{\bibfnamefont{M.}~\bibnamefont{{\v{Z}}nidari{\v{c}}}},
  \bibinfo{author}{\bibfnamefont{A.}~\bibnamefont{Scardicchio}},
  \bibnamefont{and} \bibinfo{author}{\bibfnamefont{V.~K.} \bibnamefont{Varma}},
  \bibinfo{journal}{Phys. Rev. Lett.} \textbf{\bibinfo{volume}{117}},
  \bibinfo{pages}{040601} (\bibinfo{year}{2016}).

\bibitem[{\citenamefont{Schulz et~al.}(2018)\citenamefont{Schulz, Taylor,
  Hooley, and Scardicchio}}]{schulz2018energy}
\bibinfo{author}{\bibfnamefont{M.}~\bibnamefont{Schulz}},
  \bibinfo{author}{\bibfnamefont{S.~R.} \bibnamefont{Taylor}},
  \bibinfo{author}{\bibfnamefont{C.~A.} \bibnamefont{Hooley}},
  \bibnamefont{and}
  \bibinfo{author}{\bibfnamefont{A.}~\bibnamefont{Scardicchio}},
  \bibinfo{journal}{Phys. Rev. B} \textbf{\bibinfo{volume}{98}},
  \bibinfo{pages}{180201} (\bibinfo{year}{2018}).

\bibitem[{\citenamefont{Mendoza-Arenas
  et~al.}(2019)\citenamefont{Mendoza-Arenas, {\v{Z}}nidari{\v{c}}, Varma,
  Goold, Clark, and Scardicchio}}]{mendoza2019asymmetry}
\bibinfo{author}{\bibfnamefont{J.~J.} \bibnamefont{Mendoza-Arenas}},
  \bibinfo{author}{\bibfnamefont{M.}~\bibnamefont{{\v{Z}}nidari{\v{c}}}},
  \bibinfo{author}{\bibfnamefont{V.~K.} \bibnamefont{Varma}},
  \bibinfo{author}{\bibfnamefont{J.}~\bibnamefont{Goold}},
  \bibinfo{author}{\bibfnamefont{S.~R.} \bibnamefont{Clark}}, \bibnamefont{and}
  \bibinfo{author}{\bibfnamefont{A.}~\bibnamefont{Scardicchio}},
  \bibinfo{journal}{Phys. Rev. B} \textbf{\bibinfo{volume}{99}},
  \bibinfo{pages}{094435} (\bibinfo{year}{2019}).

\bibitem[{\citenamefont{Lucioni et~al.}(2011)\citenamefont{Lucioni, Deissler,
  Tanzi, Roati, Zaccanti, Modugno, Larcher, Dalfovo, Inguscio, and
  Modugno}}]{lucioni2011observation}
\bibinfo{author}{\bibfnamefont{E.}~\bibnamefont{Lucioni}},
  \bibinfo{author}{\bibfnamefont{B.}~\bibnamefont{Deissler}},
  \bibinfo{author}{\bibfnamefont{L.}~\bibnamefont{Tanzi}},
  \bibinfo{author}{\bibfnamefont{G.}~\bibnamefont{Roati}},
  \bibinfo{author}{\bibfnamefont{M.}~\bibnamefont{Zaccanti}},
  \bibinfo{author}{\bibfnamefont{M.}~\bibnamefont{Modugno}},
  \bibinfo{author}{\bibfnamefont{M.}~\bibnamefont{Larcher}},
  \bibinfo{author}{\bibfnamefont{F.}~\bibnamefont{Dalfovo}},
  \bibinfo{author}{\bibfnamefont{M.}~\bibnamefont{Inguscio}}, \bibnamefont{and}
  \bibinfo{author}{\bibfnamefont{G.}~\bibnamefont{Modugno}},
  \bibinfo{journal}{Phys. Rev. Lett.} \textbf{\bibinfo{volume}{106}},
  \bibinfo{pages}{230403} (\bibinfo{year}{2011}).

\bibitem[{\citenamefont{Crespi et~al.}(2013)\citenamefont{Crespi, Osellame,
  Ramponi, Giovannetti, Fazio, Sansoni, De~Nicola, Sciarrino, and
  Mataloni}}]{crespi2013anderson}
\bibinfo{author}{\bibfnamefont{A.}~\bibnamefont{Crespi}},
  \bibinfo{author}{\bibfnamefont{R.}~\bibnamefont{Osellame}},
  \bibinfo{author}{\bibfnamefont{R.}~\bibnamefont{Ramponi}},
  \bibinfo{author}{\bibfnamefont{V.}~\bibnamefont{Giovannetti}},
  \bibinfo{author}{\bibfnamefont{R.}~\bibnamefont{Fazio}},
  \bibinfo{author}{\bibfnamefont{L.}~\bibnamefont{Sansoni}},
  \bibinfo{author}{\bibfnamefont{F.}~\bibnamefont{De~Nicola}},
  \bibinfo{author}{\bibfnamefont{F.}~\bibnamefont{Sciarrino}},
  \bibnamefont{and} \bibinfo{author}{\bibfnamefont{P.}~\bibnamefont{Mataloni}},
  \bibinfo{journal}{Nat. Photon.} \textbf{\bibinfo{volume}{7}},
  \bibinfo{pages}{322} (\bibinfo{year}{2013}).

\bibitem[{\citenamefont{De~Nicola et~al.}(2014)\citenamefont{De~Nicola,
  Sansoni, Crespi, Ramponi, Osellame, Giovannetti, Fazio, Mataloni, and
  Sciarrino}}]{de2014quantum}
\bibinfo{author}{\bibfnamefont{F.}~\bibnamefont{De~Nicola}},
  \bibinfo{author}{\bibfnamefont{L.}~\bibnamefont{Sansoni}},
  \bibinfo{author}{\bibfnamefont{A.}~\bibnamefont{Crespi}},
  \bibinfo{author}{\bibfnamefont{R.}~\bibnamefont{Ramponi}},
  \bibinfo{author}{\bibfnamefont{R.}~\bibnamefont{Osellame}},
  \bibinfo{author}{\bibfnamefont{V.}~\bibnamefont{Giovannetti}},
  \bibinfo{author}{\bibfnamefont{R.}~\bibnamefont{Fazio}},
  \bibinfo{author}{\bibfnamefont{P.}~\bibnamefont{Mataloni}}, \bibnamefont{and}
  \bibinfo{author}{\bibfnamefont{F.}~\bibnamefont{Sciarrino}},
  \bibinfo{journal}{Phys. Rev. A} \textbf{\bibinfo{volume}{89}},
  \bibinfo{pages}{032322} (\bibinfo{year}{2014}).

\bibitem[{\citenamefont{Geraldi et~al.}(2019)\citenamefont{Geraldi, Laneve,
  Bonavena, Sansoni, Ferraz, Fratalocchi, Sciarrino, Cuevas, and
  Mataloni}}]{geraldi2019experimental}
\bibinfo{author}{\bibfnamefont{A.}~\bibnamefont{Geraldi}},
  \bibinfo{author}{\bibfnamefont{A.}~\bibnamefont{Laneve}},
  \bibinfo{author}{\bibfnamefont{L.~D.} \bibnamefont{Bonavena}},
  \bibinfo{author}{\bibfnamefont{L.}~\bibnamefont{Sansoni}},
  \bibinfo{author}{\bibfnamefont{J.}~\bibnamefont{Ferraz}},
  \bibinfo{author}{\bibfnamefont{A.}~\bibnamefont{Fratalocchi}},
  \bibinfo{author}{\bibfnamefont{F.}~\bibnamefont{Sciarrino}},
  \bibinfo{author}{\bibfnamefont{{\'A}.}~\bibnamefont{Cuevas}},
  \bibnamefont{and} \bibinfo{author}{\bibfnamefont{P.}~\bibnamefont{Mataloni}},
  \bibinfo{journal}{Phys. Rev. Lett.} \textbf{\bibinfo{volume}{123}},
  \bibinfo{pages}{140501} (\bibinfo{year}{2019}).

\bibitem[{\citenamefont{Schreiber et~al.}(2011)\citenamefont{Schreiber,
  Cassemiro, Poto{\v{c}}ek, G{\'a}bris, Jex, and
  Silberhorn}}]{schreiber2011decoherence}
\bibinfo{author}{\bibfnamefont{A.}~\bibnamefont{Schreiber}},
  \bibinfo{author}{\bibfnamefont{K.}~\bibnamefont{Cassemiro}},
  \bibinfo{author}{\bibfnamefont{V.}~\bibnamefont{Poto{\v{c}}ek}},
  \bibinfo{author}{\bibfnamefont{A.}~\bibnamefont{G{\'a}bris}},
  \bibinfo{author}{\bibfnamefont{I.}~\bibnamefont{Jex}}, \bibnamefont{and}
  \bibinfo{author}{\bibfnamefont{C.}~\bibnamefont{Silberhorn}},
  \bibinfo{journal}{Phys. Rev. Lett.} \textbf{\bibinfo{volume}{106}},
  \bibinfo{pages}{180403} (\bibinfo{year}{2011}).

\bibitem[{\citenamefont{Anderson}(1958)}]{Anderson1958localization}
\bibinfo{author}{\bibfnamefont{P.~W.} \bibnamefont{Anderson}},
  \bibinfo{journal}{Phys. Rev.} \textbf{\bibinfo{volume}{109}},
  \bibinfo{pages}{1492} (\bibinfo{year}{1958}).

\bibitem[{\citenamefont{Geraldi et~al.}(2020)\citenamefont{Geraldi, De, Laneve,
  Barkhofen, Sperling, Mataloni, and Silberhorn}}]{geraldi2020subdiffusion}
\bibinfo{author}{\bibfnamefont{A.}~\bibnamefont{Geraldi}},
  \bibinfo{author}{\bibfnamefont{S.}~\bibnamefont{De}},
  \bibinfo{author}{\bibfnamefont{A.}~\bibnamefont{Laneve}},
  \bibinfo{author}{\bibfnamefont{S.}~\bibnamefont{Barkhofen}},
  \bibinfo{author}{\bibfnamefont{J.}~\bibnamefont{Sperling}},
  \bibinfo{author}{\bibfnamefont{P.}~\bibnamefont{Mataloni}}, \bibnamefont{and}
  \bibinfo{author}{\bibfnamefont{C.}~\bibnamefont{Silberhorn}},
  \bibinfo{journal}{arXiv:2007.12526 [quant-ph]}  (\bibinfo{year}{2020}).

\bibitem[{\citenamefont{Vakulchyk et~al.}(2019)\citenamefont{Vakulchyk, Fistul,
  and Flach}}]{vakulchyk2019wave}
\bibinfo{author}{\bibfnamefont{I.}~\bibnamefont{Vakulchyk}},
  \bibinfo{author}{\bibfnamefont{M.~V.} \bibnamefont{Fistul}},
  \bibnamefont{and} \bibinfo{author}{\bibfnamefont{S.}~\bibnamefont{Flach}},
  \bibinfo{journal}{Phys. Rev. Lett.} \textbf{\bibinfo{volume}{122}},
  \bibinfo{pages}{040501} (\bibinfo{year}{2019}).

\bibitem[{\citenamefont{Giovannetti et~al.}(2011)\citenamefont{Giovannetti,
  Lloyd, and Maccone}}]{giovannetti2011advances}
\bibinfo{author}{\bibfnamefont{V.}~\bibnamefont{Giovannetti}},
  \bibinfo{author}{\bibfnamefont{S.}~\bibnamefont{Lloyd}}, \bibnamefont{and}
  \bibinfo{author}{\bibfnamefont{L.}~\bibnamefont{Maccone}},
  \bibinfo{journal}{Nat. Photon.} \textbf{\bibinfo{volume}{5}},
  \bibinfo{pages}{222} (\bibinfo{year}{2011}).

\bibitem[{\citenamefont{Ahlbrecht et~al.}(2011)\citenamefont{Ahlbrecht, Vogts,
  Werner, and Werner}}]{ahlbrecht2011asymptotic}
\bibinfo{author}{\bibfnamefont{A.}~\bibnamefont{Ahlbrecht}},
  \bibinfo{author}{\bibfnamefont{H.}~\bibnamefont{Vogts}},
  \bibinfo{author}{\bibfnamefont{A.~H.} \bibnamefont{Werner}},
  \bibnamefont{and} \bibinfo{author}{\bibfnamefont{R.~F.}
  \bibnamefont{Werner}}, \bibinfo{journal}{J. Math. Phys.}
  \textbf{\bibinfo{volume}{52}}, \bibinfo{pages}{042201}
  (\bibinfo{year}{2011}).

\bibitem[{\citenamefont{Helstrom}(1976)}]{helstrom1976quantum}
\bibinfo{author}{\bibfnamefont{C.~W.} \bibnamefont{Helstrom}},
  \emph{\bibinfo{title}{Quantum detection and estimation theory}},
  vol.~\bibinfo{volume}{3} (\bibinfo{publisher}{Academic Press, New York},
  \bibinfo{year}{1976}).

\bibitem[{\citenamefont{Holevo}(2011)}]{holevo2011probabilistic}
\bibinfo{author}{\bibfnamefont{A.~S.} \bibnamefont{Holevo}},
  \emph{\bibinfo{title}{Probabilistic and statistical aspects of quantum
  theory}}, vol.~\bibinfo{volume}{1} (\bibinfo{publisher}{Springer Science \&
  Business Media}, \bibinfo{year}{2011}).

\bibitem[{\citenamefont{Petz and Sud{\'a}r}(1996)}]{petz1996geometries}
\bibinfo{author}{\bibfnamefont{D.}~\bibnamefont{Petz}} \bibnamefont{and}
  \bibinfo{author}{\bibfnamefont{C.}~\bibnamefont{Sud{\'a}r}},
  \bibinfo{journal}{J. Math. Phys.} \textbf{\bibinfo{volume}{37}},
  \bibinfo{pages}{2662} (\bibinfo{year}{1996}).

\bibitem[{\citenamefont{Petz}(2002)}]{petz2002covariance}
\bibinfo{author}{\bibfnamefont{D.}~\bibnamefont{Petz}}, \bibinfo{journal}{J.
  Phys. A: Math. Gen.} \textbf{\bibinfo{volume}{35}}, \bibinfo{pages}{929}
  (\bibinfo{year}{2002}).

\bibitem[{\citenamefont{Paris}(2009)}]{paris2009quantum}
\bibinfo{author}{\bibfnamefont{M.~G.~A.} \bibnamefont{Paris}},
  \bibinfo{journal}{Int. J. Quant. Inf.} \textbf{\bibinfo{volume}{7}},
  \bibinfo{pages}{125} (\bibinfo{year}{2009}).

\bibitem[{\citenamefont{{Lo Franco} and Compagno}(2016)}]{lofranco2016}
\bibinfo{author}{\bibfnamefont{R.}~\bibnamefont{{Lo Franco}}} \bibnamefont{and}
  \bibinfo{author}{\bibfnamefont{G.}~\bibnamefont{Compagno}},
  \bibinfo{journal}{Sci. Rep.} \textbf{\bibinfo{volume}{6}},
  \bibinfo{pages}{20603} (\bibinfo{year}{2016}).

\bibitem[{\citenamefont{Killoran et~al.}(2014)\citenamefont{Killoran, Cramer,
  and Plenio}}]{plenio2014PRL}
\bibinfo{author}{\bibfnamefont{N.}~\bibnamefont{Killoran}},
  \bibinfo{author}{\bibfnamefont{M.}~\bibnamefont{Cramer}}, \bibnamefont{and}
  \bibinfo{author}{\bibfnamefont{M.~B.} \bibnamefont{Plenio}},
  \bibinfo{journal}{Phys. Rev. Lett.} \textbf{\bibinfo{volume}{112}},
  \bibinfo{pages}{150501} (\bibinfo{year}{2014}).

\bibitem[{\citenamefont{Compagno et~al.}(2018)\citenamefont{Compagno,
  Castellini, and {Lo Franco}}}]{compagno2018}
\bibinfo{author}{\bibfnamefont{G.}~\bibnamefont{Compagno}},
  \bibinfo{author}{\bibfnamefont{A.}~\bibnamefont{Castellini}},
  \bibnamefont{and} \bibinfo{author}{\bibfnamefont{R.}~\bibnamefont{{Lo
  Franco}}}, \bibinfo{journal}{Phil. Trans. R. Soc. A}
  \textbf{\bibinfo{volume}{376}}, \bibinfo{pages}{20170317}
  (\bibinfo{year}{2018}).

\bibitem[{\citenamefont{{Lo Franco} and Compagno}(2018)}]{lofranco2018PRL}
\bibinfo{author}{\bibfnamefont{R.}~\bibnamefont{{Lo Franco}}} \bibnamefont{and}
  \bibinfo{author}{\bibfnamefont{G.}~\bibnamefont{Compagno}},
  \bibinfo{journal}{Phys. Rev. Lett.} \textbf{\bibinfo{volume}{120}},
  \bibinfo{pages}{240403} (\bibinfo{year}{2018}).

\bibitem[{\citenamefont{Nosrati et~al.}(2020)\citenamefont{Nosrati, Castellini,
  Compagno, and {Lo Franco}}}]{nosrati2019npj}
\bibinfo{author}{\bibfnamefont{F.}~\bibnamefont{Nosrati}},
  \bibinfo{author}{\bibfnamefont{A.}~\bibnamefont{Castellini}},
  \bibinfo{author}{\bibfnamefont{G.}~\bibnamefont{Compagno}}, \bibnamefont{and}
  \bibinfo{author}{\bibfnamefont{R.}~\bibnamefont{{Lo Franco}}},
  \bibinfo{journal}{npj Quantum Information} \textbf{\bibinfo{volume}{6}},
  \bibinfo{pages}{39} (\bibinfo{year}{2020}).

\bibitem[{\citenamefont{Sun et~al.}(2020)\citenamefont{Sun, Wang, Liu, Xu, Xu,
  Li, Guo, Castellini, Nosrati, Compagno et~al.}}]{sunetalexp}
\bibinfo{author}{\bibfnamefont{K.}~\bibnamefont{Sun}},
  \bibinfo{author}{\bibfnamefont{Y.}~\bibnamefont{Wang}},
  \bibinfo{author}{\bibfnamefont{Z.-H.} \bibnamefont{Liu}},
  \bibinfo{author}{\bibfnamefont{X.-Y.} \bibnamefont{Xu}},
  \bibinfo{author}{\bibfnamefont{J.-S.} \bibnamefont{Xu}},
  \bibinfo{author}{\bibfnamefont{C.-F.} \bibnamefont{Li}},
  \bibinfo{author}{\bibfnamefont{G.-C.} \bibnamefont{Guo}},
  \bibinfo{author}{\bibfnamefont{A.}~\bibnamefont{Castellini}},
  \bibinfo{author}{\bibfnamefont{F.}~\bibnamefont{Nosrati}},
  \bibinfo{author}{\bibfnamefont{G.}~\bibnamefont{Compagno}},
  \bibnamefont{et~al.}, \bibinfo{journal}{Opt. Lett., in press; Preprint at
  arXiv:2003.10659 [quant-ph]}  (\bibinfo{year}{2020}).

\end{thebibliography}


\appendix

\section{Appendix: Further analysis}

\begin{figure*}[t]
\begin{center}
\subfloat[Static disorder] {\includegraphics[width=0.45\textwidth]{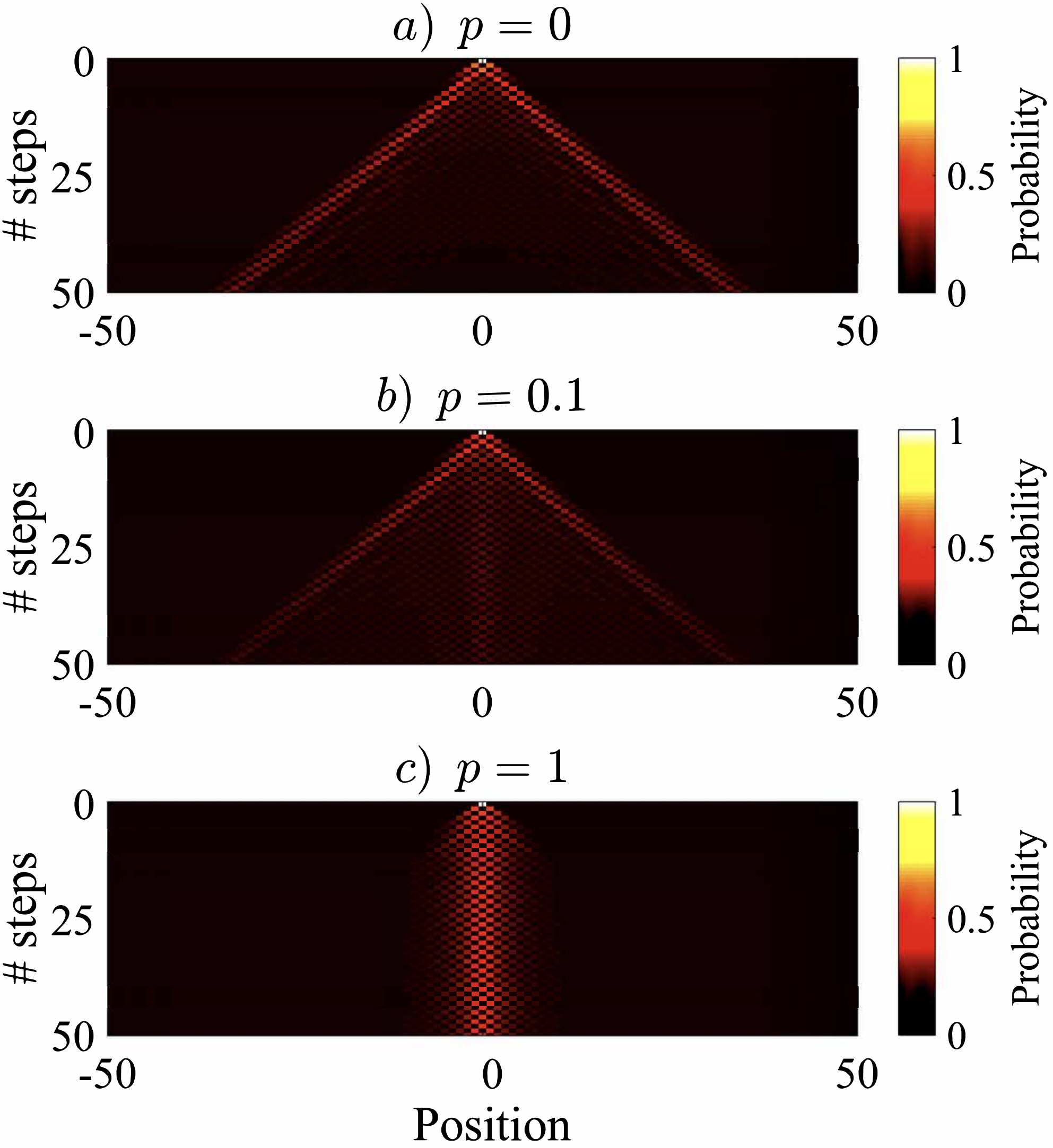}}
\subfloat[Dynamic disorder] {\includegraphics[width=0.45\textwidth]{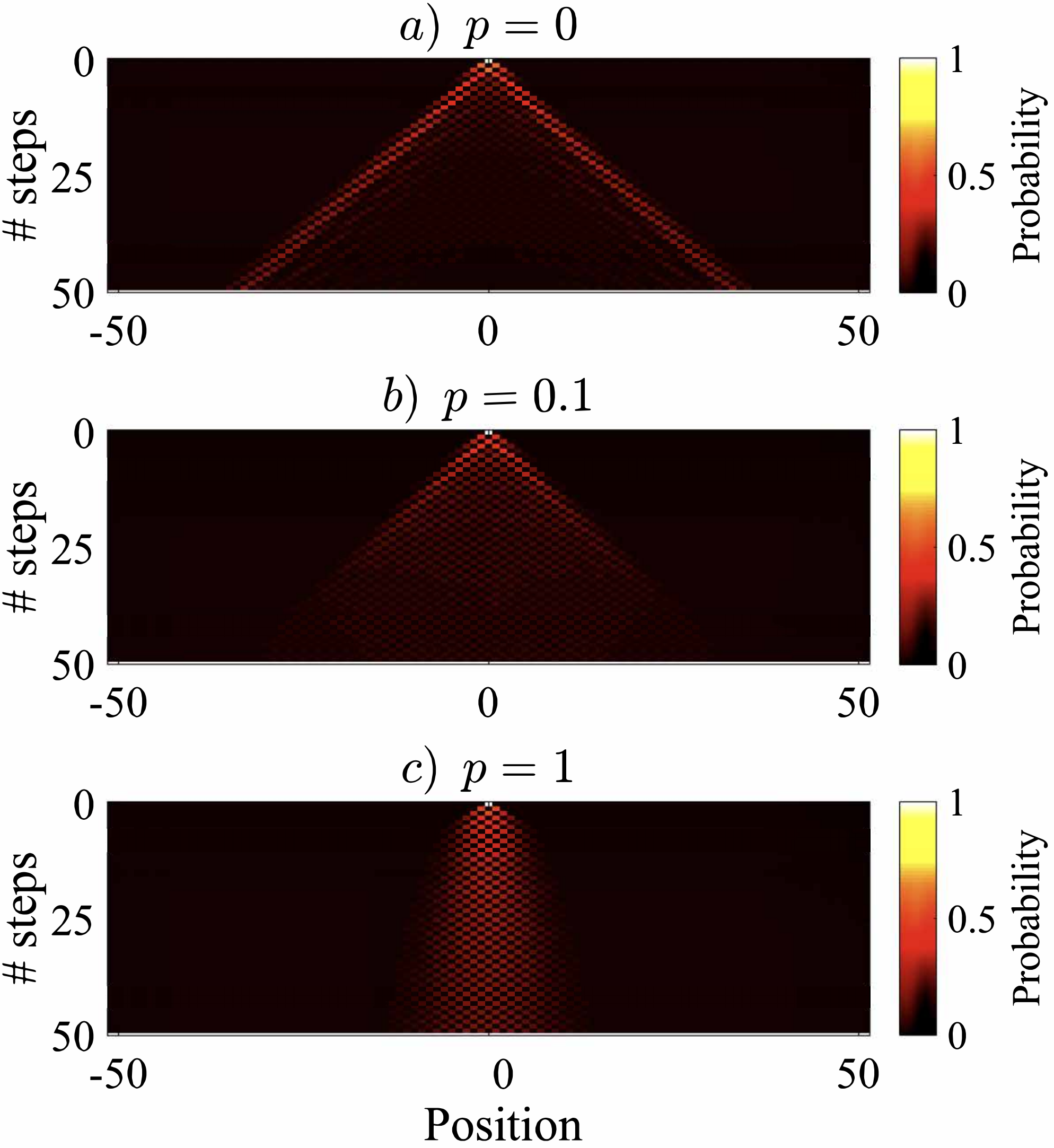}}
\end{center}
\caption{\textbf{Probability distribution.} Average probability distribution of an unbounded discrete QW on the line as a function of the step number and of the position, for different values of disorder: \textbf{a}) $p=0$ \textbf{b)}, $p=0.1$, \textbf{c)} $p=1$, by averaging over $10^4$ phase maps for both static disorder (\textbf{I}) and dynamic disorder (\textbf{II}). In all cases, the initial input state  is $\ket{\Psi_0}=\frac{1}{\sqrt{2}}\ket{0}_p\otimes\left(\ket{\uparrow}_c+\ket{\downarrow}_c\right)$}\label{prb-fig}
\end{figure*}

A typical representation of a disordered quantum walk network is displayed in Fig. \ref{fig-repe} for four steps. Each line linking two different vertical bars corresponds to a mode along which the walker can travel. In the dynamic disorder, each step walker acquires a phase shift depending on the step and the mode. Each vertical line represents the unitary operator for a certain site and step of the evolution. As a consequence, the walker state at step $t$ is given by $t$ repetitive action of step-dependent unitary operator:
\begin{equation}
    \ket{\Psi_t}=\hat{U}_t\hat{U}_{t-1}\dots\hat{U}_2\hat{U}_1\ket{\Psi_0}.
\end{equation}
Here, each unitary operator $\hat{U}_t$ is given with the specifics phase map $\hat{P}$ in the form of 
\begin{equation}
    \hat{P}=\sum_x\ket{x}_{\mathbf{p}}\bra{x}_{\mathbf{p}}\otimes\left(\ket{\downarrow}_{\mathbf{c}}\bra{\downarrow}_{\mathbf{c}}+e^{\mathbf{i}\left(\Delta\phi'(t,x)\right)}\ket{\uparrow}_{\mathbf{c}}\bra{\uparrow}_{\mathbf{c}}\right).
\end{equation}
In addition, one might consider static disorder where phase fluctuations are frozen in time, therefore the quantum state at step $t$ is $ \ket{\Psi_t}=(\hat{U})^{t}\ket{\Psi_0}$. Also, we define the degree of disorder $p$ as percentage of random phases that the walker experiences during the evolution. This means different types of phase maps can be realized for any given value of the degree of disorder.

Let us consider a given initial state 
\begin{equation}\label{Psi0Supp}
    \ket{\Psi_0}=\frac{1}{\sqrt{2}}\ket{0}_{\mathbf{p}}\otimes\left(\ket{\uparrow}_{\mathbf{c}}+\ket{\downarrow}_{\mathbf{c}}\right),
\end{equation}
to explore the probability distribution of a quantum walker in presence of both static and dynamic disorder. We limit our analysis in the case that a $p$ percentage of phase fluctuations are selected out of $0$ and $\pi$.  
We show the density plots of the evolution of single particles in Fig. \ref{prb-fig}, in terms of step number and walker position. Here, both  static (Fig. \ref{prb-fig} (\textbf{I})) and dynamic (Fig. \ref{prb-fig} (\textbf{II})) disorder is realized with different degrees of disorder $p$: \textbf{a}) $p=0$, which corresponds to a standard ordered QW; \textbf{b}) $p=0.1$; \textbf{c}) $p=1$ completely disordered QW. The horizontal axis denotes different positions that the walker can reach while the vertical one represents the step number $t$, increasing from top to bottom.  We averaged  the  probability  distributions over $10000$ random phase maps realizations. As can be seen, the particle follows different transport pattern effected by static  (Fig. \ref{prb-fig} (\textbf{I})) or dynamic disorder ((\textbf{II})). Also, we show how the degree of disorder would affect the probability distribution of the quantum walker. It seems that the probability of finding a particle in the center raises by increasing the degree of disorder $p$.

\begin{figure*}[t]
\begin{center}
\subfloat[Static disorder] {\includegraphics[width=0.95\textwidth]{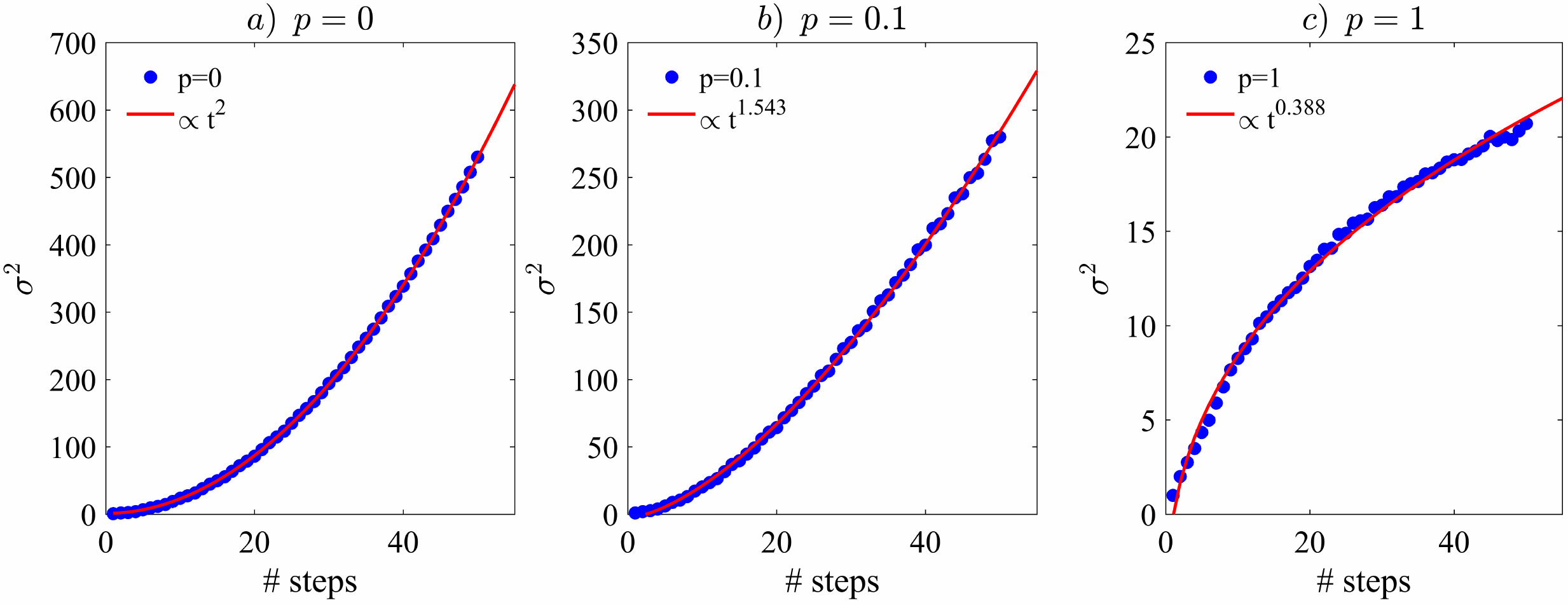}} 

\subfloat[Dynamic disorder] {\includegraphics[width=0.95\textwidth]{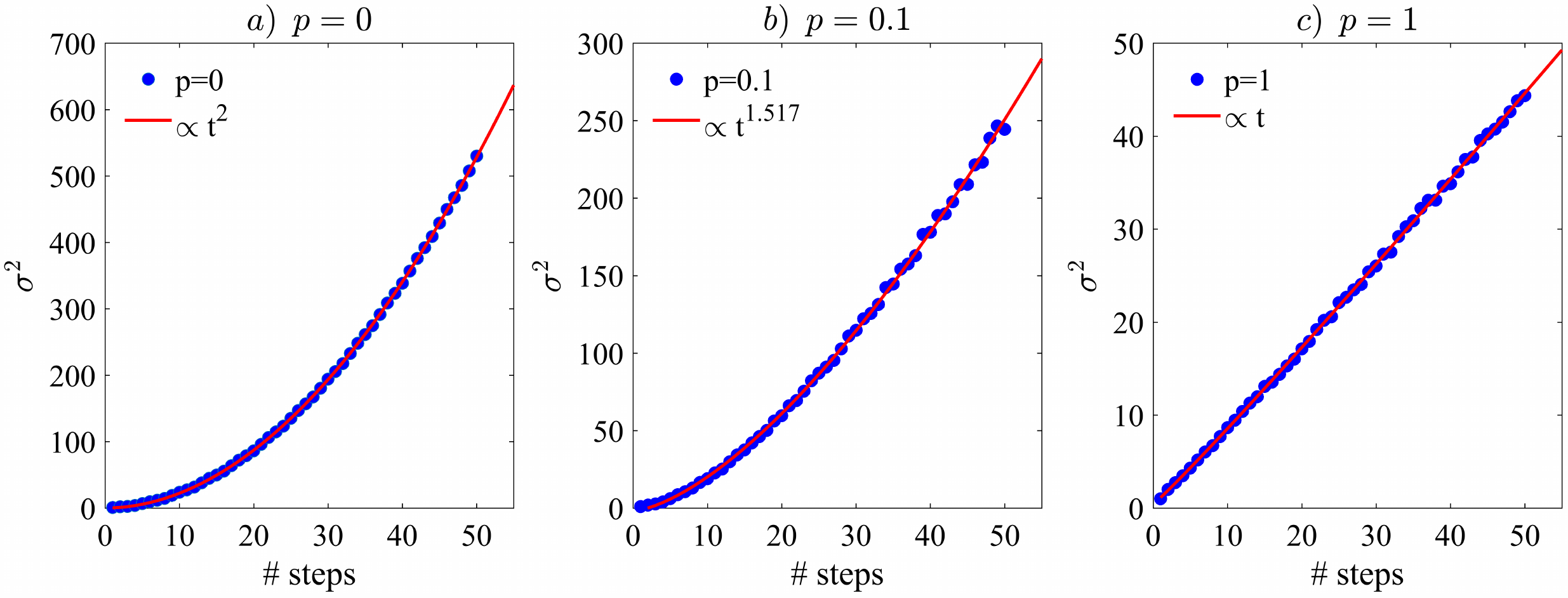}} 
\end{center}
\caption{\textbf{Position variance.} Position variance $\sigma^2$ of a quantum walker on the line in terms of the step number $t$ for different values of disorder \textbf{a}) $p=0$ \textbf{b)}, $p=0.1$, and \textbf{c)} $p=1$, by averaging over $10000$ phase maps  for both static disorder (\textbf{I}) and dynamic disorder (\textbf{II}). In all cases the initial input is $\ket{\Psi_0}=\frac{1}{\sqrt{2}}\ket{0}_p\otimes\left(\ket{\uparrow}_c+\ket{\downarrow}_c\right)$}\label{var-fig}
\end{figure*}

In the QW problem, the operator $\hat{X}$ that measures the walker position is the operator of interest since the variance of the position operator provides a simple measure to quantify the spread of the walker \cite{schreiber2011decoherence}
\begin{equation}
    \sigma^2(\hat{X})=\langle\hat{X}^2\rangle-\langle\hat{X}\rangle^2.
\end{equation}
This measure has been proven to be particularly useful to compare the effects of different kinds of disorder on the spreading pattern of the walker(s). Generally, the variance of the position operator is given by $\sigma^2(\hat{X})\propto t^\alpha$, where range $\alpha>1$ is called superdiffusive  and $\alpha<1$ subdiffusive. For instance, an ordered QW presents a ballistic spread with $\sigma^2(\hat{X})\propto t^2$, while the classical random walk is diffusive with $\sigma^2(\hat{X})\propto t$ .  

As an example, we study the spreading behavior of a walker for the given input state $\ket{\Psi_0}$ of Eq.~\eqref{Psi0Supp} with three different values of disorder. The position variance of a quantum walker generally depends on type of the disorder, degrees of disorder $p$ and number of steps $t$. The variance of a single walker is plotted versus the number of steps for both static (Fig. \ref{prb-fig} (\textbf{I})) and dynamic (Fig. \ref{prb-fig} (\textbf{II})) disorder. For an ordered quantum walk, the walker shows a ballistic pattern, displayed in Fig.\ref{var-fig} \textbf{a}. In static disorder, a super-diffusive pattern appears because of the increasing value of disorder degree to $p=0.1$ (see Fig. \ref{var-fig} (\textbf{I})\textbf{b}). Interestingly, in Fig. \ref{var-fig} (\textbf{I})\textbf{c}), the subdiffusive pattern also can be simulated by increasing the value of the disorder degree to the maximum. In dynamical disorder case, the quantum walker exhibits a transition from ballistic to superdiffusive, and then classical spreading pattern by increasing the degree of disorder to $p=1$.

\end{document}